\documentclass[preprintnumbers,amsmath,amssymb,aps,prl,groupedaddress]{revtex4}

\usepackage{multirow}
\usepackage{hyperref}
\usepackage{mathrsfs}
\usepackage{amsmath}
\usepackage{graphicx}
\usepackage{color}
\begin{document}
\draft
\title{Enhancement of sensitivity near exceptional points in dissipative qubit-resonator systems}
\author{Pei-Rong Han$^{1,2}$}
\thanks{These authors contribute equally to this work.}
\author{Fan Wu$^{2}$}
\thanks{These authors contribute equally to this work.}
\author{Xin-Jie Huang$^{2}$}
\thanks{These authors contribute equally to this work.} 
\author{Huai-Zhi Wu$^{2}$, Chang-Ling Zou$^{3,4}$, Wei Yi$^{3,4}$, Mengzhen Zhang$^{5}$, Hekang Li$^{6}$%
, Kai Xu$^{6,7}$, Dongning Zheng$^{6,7}$, Heng Fan$^{6,7}$} 
\author{Jianming Wen$^{8,9}$}
\thanks{E-mail: jianming.wen@gmail.com} 
\author{Zhen-Biao Yang$^{2}$}
\thanks{E-mail: zbyang@fzu.edu.cn}
\author{Shi-Biao Zheng$^{2}$}
\thanks{E-mail: t96034@fzu.edu.cn}
\address{$^{1}$School of Physics and Mechanical and Electrical Engineering, Longyan University, Longyan 364012, China\\
	$^{2}$Fujian Key Laboratory of Quantum Information and Quantum Optics, College of Physics and Information Engineering, Fuzhou University,\\
	Fuzhou 350108, China\\
$^{3}$CAS Key Laboratory of Quantum Information, University of Science and\\
Technology of China, Hefei 230026, China\\
$^{4}$CAS Center for Excellence in Quantum Information and Quantum Physics,\\
University of Science and Technology of China, Hefei 230026, China\\
$^{5}$Pritzker School of Molecular Engineering, University of Chicago,\\
Chicago, IL 60637, USA\\
$^{6}$Institute of Physics, Chinese Academy of Sciences, Beijing 100190,\\
China\\
$^{7}$CAS Center for Excellence in Topological Quantum Computation,\\
University of Chinese Academy of Sciences, Beijing 100190, China\\
$^{8}$Department of Electrical and Computer Engineering, Binghamton University, Binghamton, New York 13902, USA\\
$^{9}$Department of Physics, Kennesaw State University, Marietta, Georgia\\
30060, USA}
%$^{8}$Department of Physics, Kennesaw State University, Marietta, Georgia\\
%30060, USA}
%\date{\today }

\begin{abstract}
Dissipation usually plays a negative role in quantum metrological
technologies, which aim to improve measurement precision by leveraging
quantum effects that are vulnerable to environment-induced decoherence.
Recently, it has been demonstrated that dissipation can actually be used
as a favorable resource for enhancing the susceptibility of signal
detection. However, demonstrations of such enhancement for detecting
physical quantities in open quantum systems are still lacking. Here we
propose and demonstrate a protocol for realizing such non-Hermitian quantum
sensors for probing the coupling between a qubit and a resonator subjecting
to energy dissipations. The excitation-number conversion associated with the
no-jump evolution trajectory enables removal of the noisy outcomes with
quantum jumps, implementing the exceptional point (EP), where the Rabi
splitting exhibits a divergent behavior in response to a tiny variation of
the effective coupling. The sensitivity enhancement near the EP is confirmed
by both theoretical calculation and experimental measurement.
\end{abstract}

\maketitle
\vskip0.5cm

\narrowtext

\bigskip Quantum metrology represents one of the most fast-growing branches
of quantum science and technology, aiming to improve the measurement precision of physical quantities, such as the amplitude of a
magnetic or electric field, the frequency of an oscillator, and the
magnitude of a force [1-4]. The central resource of a quantum sensor is
highly nonclassical states, e.g., multi-qubit entangled states, which
exhibit strong sensitivity to variations of the control parameter.
Theoretically, the larger the scale of the quantum state, the more sensitive
the quantum sensor becomes. Practically, the quantum state becomes increasingly
vulnerable to the environment-induced decoherence with the increase in
system size, which is one of the main obstacles for the realization of quantum
techniques. For example, the advantage in sensitivity empowered by a
multi-qubit entangled state [5-9] or an \textit{N}-quanta interfering state of a bosonic
mode [10,11] is quickly compromised by the increased decoherence rate, which
limits the available sensitivity enhancement. Consequently,
proof-of-principle demonstrations of quantum advantages with such
nonclassical states have been restricted to a quite moderate level [12].

Alternative avenue for enhancing sensitivity is to exploit the exceptional
singularities associated with the non-Hermitian (NH) Hamiltonian dynamics, which
has been attracting increasing interest [13-22]. While NH physics can derive from the open quantum systems, its unique features extend the Hermitian quantum mechanics. Increasing efforts have been devoted to the exploration of NH physics, such as parity-time (PT)-symmetric theory [23,24], pseudo-Hermitian systems [25], quantum simulation of NH systems [26-28], NH entropies [29], open quantum systems [30,31], etc. In particular, the non-Hermitian 
degeneracy point, referred to as the exceptional point (EP), exhibits exotic
behaviors that are fundamentally distinct from its Hermitian counterparts,
exemplified by the coalescence of both the eigenenergies and eigenstates.
The sensitivity enhancement of an NH sensor is enabled by the divergent
derivative of the eigenenergy gap to the control parameters around the EP,
with the susceptibility enhancement dependent on the order of the EP. For
example, at a second\ order EP, the gap exhibits a square-root scaling,
such that the energy splitting around the EP becomes extremely sensitive
to a tiny change of the control parameter. EP-enhanced sensing protocols
have been demonstrated in classical optical [19,20] and electromechanical
[21] systems with artificially engineered gain-and-loss channels, and, in a recent experiment, 
using single photons [22], though no
specific physical quantity was estimated.

We here propose a theoretical scheme and present an experimental
demonstration for the NH-Hamiltonian-based quantum sensing to detect the coupling strength between a qubit and a lossy resonator [32]. Near
the EP, a slight variation of such a coupling strength results in a
response of the vacuum Rabi splitting of the coupled qubit-resonator system,
which underlies the signal amplification. For the no-jump trajectory, the
system excitation number is conserved, which enables us to infer the vacuum
Rabi oscillation signals governed by the NH Hamiltonian, by correlating the
measured output state of the qubit with that of the resonator. The vacuum
Rabi splitting, extracted from the Rabi signals associated with the no-jump
trajectory, exhibits a divergent behavior at the EP. The experimental
results confirm the non-Hermiticity-enabled sensitivity enhancement.

The NH system under consideration involves a qubit with an energy
dissipation rate of $\kappa _{q}$ and a lossy resonator with a photonic
decaying rate, denoted as $\kappa _{p}$. The qubit is resonantly coupled to
the resonator with the strength $\Omega $. In the interaction picture, the
system dynamics is described by the master equation (setting $\hbar =1$)
\begin{equation}
\frac{d\rho }{dt}=-i(H_{S}\rho -\rho H_{S}^{\dagger})+\kappa _{q}\sigma ^{-}\rho \sigma
^{+}+\kappa _{p}a\rho a^{\dagger }, 
\end{equation}
where
\begin{equation}
H_{S}=\Omega (a^{\dagger }\sigma ^{-}+a\sigma ^{+})-\frac{i}{2}\kappa
_{q}\sigma ^{+}\sigma ^{-}-\frac{i}{2}\kappa _{p}a^{\dagger }a. 
\end{equation}
Here $\sigma ^{+}=\left\vert e\right\rangle \left\langle g\right\vert $ and $%
\sigma ^{-}=\left\vert g\right\rangle \left\langle e\right\vert $ with $%
\left\vert g\right\rangle $ ($\left\vert e\right\rangle $) denoting the
lower (upper) level of the qubit, and $a^{\dagger }$\ ($a$) represents the
photonic creation (annihilation) operator for the resonator. Under the
master equation, the system evolution is a mixture of infinitely many
trajectories. The dynamics associated with the no-jump trajectory is governed
by the NH Hamiltonian $H_{S}$. We here focus on the system dynamics within
the single-excitation subspace $\{\left\vert e,0\right\rangle ,\left\vert
g,1\right\rangle \}$, where the number in each ket denotes the photon number
of the resonator. In such a subspace, the system has two entangled
eigenstates, given by
\begin{equation}
\left\vert \Phi _{\pm }\right\rangle ={\cal N}_{\pm }\left[\Omega \left\vert
e,0\right\rangle +(-i\kappa/4\pm E)\left\vert g,1\right\rangle \right], 
\end{equation}
where $E=\sqrt{\Omega ^{2}-\kappa ^{2}/16}$ is half of the vacuum Rabi
splitting between the two eigenstates, with $\kappa =\kappa _{p}-\kappa _{q}$.

\begin{figure*}[htbp]
	\centering
	\includegraphics[width=4in]{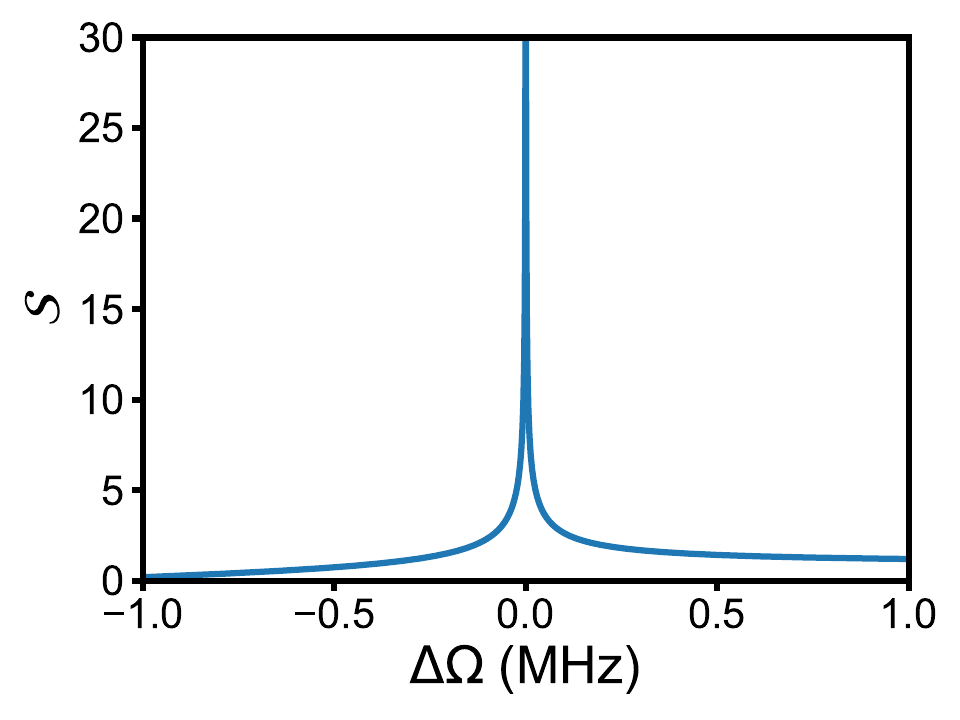}
	\caption{(color online). Calculated non-Hermiticity-enabled sensitivity
		enhancement. The qubit-resonator coupling strength $\Omega $ is inferred
		from the measured $E$, so that the signal amplification is characterized by $%
		S=\left\vert dE/d\Omega \right\vert $.}
	\label{Fig1}
\end{figure*}

The derivative of the vacuum Rabi splitting to $\Omega $ is 
\begin{equation}
dE/d\Omega =\Omega /\sqrt{\Omega ^{2}-\kappa ^{2}/16}, 
\end{equation}
which diverges at the EP $\Omega _{0}=\left\vert \kappa /4\right\vert $.
This implies that even a small variation of $\Omega $ can cause a
significant change of the Rabi splitting around the EP. This divergent
behavior is a valuable resource for estimating a slight deviation of $\Omega 
$ from $\Omega _{0}$, defined as $\Delta \Omega =\Omega -\Omega _{0}$. Above
the EP, both $E$ and $dE/d\Omega $ are real, which become imaginary below
the EP. To estimate the performance of the NH quantum sensor above and
below the EP in a unified way, we characterize the non-Hermiticity-enabled
sensitivity enhancement by $S=\left\vert dE/d\Omega \right\vert $ [20]. Figure
1 displays $S$ as a function of $\Delta
\Omega$, which shows that the closer the control parameter to the EP, the higher
the sensitivity.

\begin{figure*}[htbp]
	\centering
	\includegraphics[width=3in]{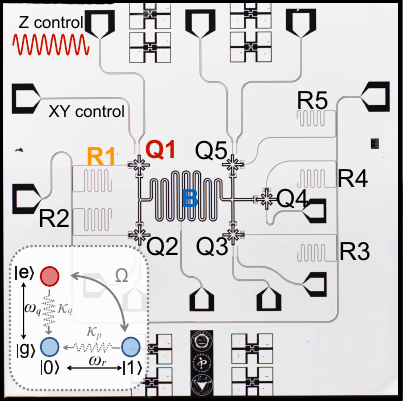}
	\caption{(color online). Experimental device. The experiment is performed
		with a circuit quantum electrodynamics architecture involving five
		frequency-tunable superconducting qubits, labeled from $Q_{1}$ to $Q_{5}$,
		each of which is individually connected to a readout resonator ($R_{j}$).
		These qubits are connected by a bus resonator ($B$). $Q_{1}$ is effectively
		coupled to $R_{1}$ by a parametric modulation, which mediates a sideband
		interaction. For the no-jump evolution trajectory, the system dynamics is
		governed by the NH Hamiltonian of Eq. (2) with $\kappa _{q}=0.07$ MHz and $\kappa
		_{p}=5$ MHz. The effective coupling strength $\Omega $ is controlled by the
		modulating amplitude. After the NH dynamics, the output $Q_{1}$-$R_{1}$
		state is mapped to $Q_{2}$-$Q_{1}$ system with the help of $B$.}
	\label{Fig2}
\end{figure*}

We synthesize such an NH model in a circuit
quantum electrodynamics architecture [32], where a bus resonator ($B$) with a fixed
frequency $\omega _{b}/2\pi=5.58$ GHz is connected to five frequency-tunable
superconducting qubits, labeled from $Q_{1}$ to $Q_{5}$, each of which is
individually connected to a readout resonator ($R_{j}$). 
In this experiment implementation, we only use two qubits ($Q_{1}$ and $Q_{2}$, while other qubits are tuned far from their sweet point to avoid unwanted parametric-modulation-induced qubit-qubit interactions), the bus resonator ($B$) and a readout resonator ($R_{1}$).
The device is
sketched in Fig. 2, where $Q_{1}$ and $R_{1}$ are used to realize the
qubit-resonator system. Their effective interaction is mediated by applying
a parametric modulation to $Q_{1}$, modulating its frequency as $\omega
_{q}=\omega _{0}+\varepsilon \cos (\nu t)$, where $\omega _{0}$ is the
mean $\left\vert e\right\rangle $-$\left\vert g\right\rangle $ energy
difference, and $\varepsilon $ and $\nu $ denote the modulating amplitude
and frequency. With suitable choice of the modulating frequency $\nu $, $%
Q_{1}$ resonantly interacts with $R_{1}$ at one sideband with respect to the
modulation. This effectively realizes the NH Hamiltonian of Eq. (2), where $%
\kappa _{q}=0.07$ MHz, $\kappa _{p}=5$ MHz, and the effective coupling strength is
controllable by the modulating amplitude $\varepsilon $.

\begin{figure*}[htbp]
	\centering
	\includegraphics[width=4in]{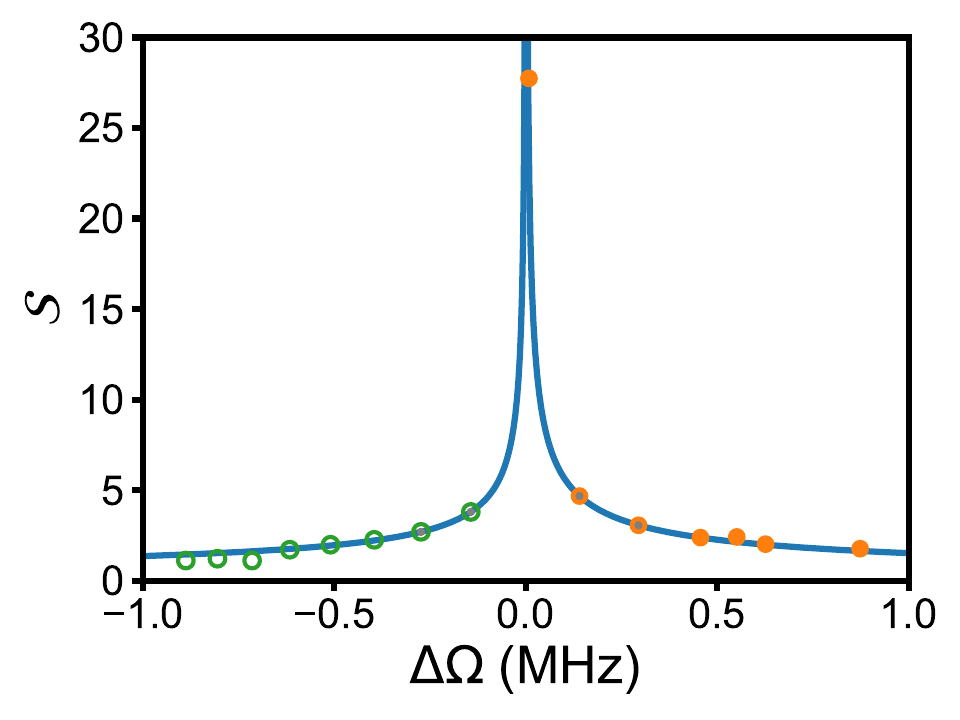}
	\caption{(color online). Demonstration of sensitivity enhancement. The data at
		each point is related to the measured $E$ by $S=(\mathop{\rm Re}E-\mathop{\rm Im}E)/\Delta\Omega$, 
		where $\Delta\Omega$ denotes the deviation of the effective $Q_{1}$-$R_{1}$ coupling from the
		EP. The lines are the power law fitting $S=A\left\vert \Delta\Omega/\Omega_{0}\right\vert ^{B}$.}
	\label{Fig3}
\end{figure*}

The $Q_{1}$-$R_{1}$ system is initially prepared in the state $\left\vert
e,0\right\rangle $. Then the sideband interaction is initiated by the
application of the modulating flux, which evolves the system in the
single-excitation subspace $\{\left\vert g,1\right\rangle ,\left\vert
e,0\right\rangle \}$. After a preset interaction time $t$, the modulation is
switched off, following which the $Q_{1}$-$R_{1}$ state is read out. 
The state of $R_{1}$ cannot be read out directly. To measure the state of $R_{1}$, state tansfer and mapping operations are implemented. This is achieved by transferring the state of $Q_{1}$ to an
ancilla qubit ($Q_{2}$) through $B$, realized with swapping gates, and then
performing the $R_{1}\rightarrow Q_{1}$ mapping operation. With the state
distortion due to $R$'s dissipation during the state transfer being corrected
for, the resulting joint $Q_{2}$-$Q_{1}$ state corresponds to the $Q_{1}$-$%
R_{1}$ output state. The state evolution trajectory governed by the NH
Hamiltonian is obtained by discarding the measurement outcome $\left\vert
g,g\right\rangle $, and renormalizing the remaining single-excitation
populations.  
The eigenenergies $E$ are obtained by measuring the time-evolving output states associated with the no-jump trajectory. As the state evolution can be expanded by the eigenstates and eigenenergies of the NH Hamiltonian, we apply the least-squares fitting to the measured density matrix to extract both the real and imaginary parts of the eigenenergies. The fitting results display a real-to-imaginary transition at the EP, as shown in Fig. 3(d) of Ref. [32].
We here use the data measured in Ref. [32] to demonstrate
the EP-enhanced sensitivity for estimating $\Omega $.

The divergent behavior of $E$ implies that it serves as a sensitive
indicator for estimating a slight deviation of $\Omega$ from $\Omega_{0}$, 
i.e., $\Delta\Omega$. Figure 3 displays $S$ as a function of $
\Omega$, where the data at each point is related to the measured $E$ by 
$S=(\mathop{\rm Re}E-\mathop{\rm Im}E)/\Delta\Omega$. 
The lines, which denote the functional form of $S$ in the vicinity of the EP, are obtained by
fitting the experimental data to the power law function 
$S=A\left\vert\Delta\Omega/\Omega_{0}\right\vert ^{B}$. 
The fitted parameter values for $\Delta\Omega>0$ are $A=1.349$ and $B=-0.572$, 
while for $\Delta\Omega<0$ they are $A=1.215$  and $B=-0.530$. The results unambiguously confirm that the EP
serves as a favorable resource for enhanced sensing. We note that the
signal amplification is enabled by the naturally-occurring
dissipations, which represent one of the main drawbacks restricting
the precision for estimating inherent properties of open quantum systems
[1]. Our qubit-resonator system offers the possibility to turn the
resulting dissipation from an unavoidable nuisance into a stable resource
for enhancing sensitivity.  This feature originates from the inherent
excitation conservation associated with the no-jump trajectory, which enables the
removal of the resulting noisy data from the useful results.

In conclusion, we have proposed and demonstrated an NH quantum
sensing protocol for estimating the coupling strength between a qubit and a
resonator. Around the EP of the strongly interacting light-matter system,
the signal is amplified due to the square-root dependence of the vacuum Rabi
splitting on the deviations of the control parameters from the EP values.
The vacuum Rabi splitting is extracted from the measured population
evolution in the subspace governed by the NH Hamiltonian, thanks to the U(1)
symmetry of the system, which enables the noisy outcomes to be distinguished
from the no-jump results. The experimental results confirm that the
sensitivity is significantly enhanced by the EP. The method can
be directly generalized to estimate the coupling strength between two
interacting qubits with unbalanced decaying rates.

This work was supported by the
National Natural Science Foundation of China (Grants
No. 12274080, No. 12474356, No. 12475015, No. 12204105,
No. 11774058, No. 12174058, No. 12374479,
No. 11934018, No. 92065114, and No. T2121001),
Innovation Program for Quantum Science and Technology
(Grants No. 2021ZD0300200 and No. 2021ZD0301200),
NSF (Grant No. 2329027), the Strategic Priority Research
Program of Chinese Academy of Sciences (Grant
No. XDB28000000), the National Key R\&D Program under
(Grant No. 2017YFA0304100), the Key-Area Research
and Development Program of Guangdong Province, China
(Grant No. 2020B0303030001), Beijing Natural Science
Foundation (Grant No. Z200009), and the Project from
Fuzhou University (Grant No. 049050011050).

%\clearpage

\end{document}